\documentclass[pra,aps,floatfix,twocolumn]{revtex4}
\usepackage{makeidx,epsfig}
\usepackage{setspace,graphicx}%,subfigure
\usepackage{amsmath,dcolumn}

\begin{document}

\newcommand{\etal}{{\it et al.}}
\newcommand{\fref}[1]{Fig.~\ref{#1}}
\newcommand{\Fref}[1]{Figure \ref{#1}}
\newcommand{\sref}[1]{Sec. \ref{#1}}
\newcommand{\Eref}[1]{Eq.~(\ref{#1})}
\newcommand{\tref}[1]{Table~\ref{#1}}
\newcommand{\rtw}{\rightarrow}
\def\veps{\varepsilon}

\title{Studying variation of fundamental constants with molecules}
\author{V. V. Flambaum}
\affiliation{School of Physics, The University of New South Wales,
Sydney NSW 2052, Australia}
\author{M. G. Kozlov}
\affiliation{Petersburg Nuclear Physics Institute, Gatchina 188300,
             Russia}
\maketitle

\section{Introduction}\label{intro}

In this chapter we will discuss an application of precision
molecular spectroscopy to the studies of the possible spatial and
temporal variations of the fundamental constants. As we will see
below, molecular spectra are mostly sensitive to two such
dimensionless constants, namely the fine-structure constant
$\alpha=\tfrac{e^2}{\hbar c}$ and the electron-to-proton mass ratio
$\mu=m_e/m_p$ (note that some papers define $\mu$ as an inverse
value, i.e. proton-to-electron mass ratio). At present NIST gives
following values of these constants~\cite{NIST}:
$\alpha^{-1}=137.035999679(94)$ and $\mu^{-1}=1836.15267247(80)$.

The fine-structure constant $\alpha$ determines the strength of
electromagnetic (and more generally electroweak) interactions. In
principle, there is similar coupling constant $\alpha_s$ for quantum
chromodynamics (QCD). However, because of the highly nonlinear
character of the strong interactions, this constant is not well
defined. Therefore, the strength of the strong interactions is
usually characterized by the parameter $\Lambda_\mathrm{QCD}$, which
has the dimension of mass and is defined as the position of the
Landau pole in the logarithm for the running strong coupling
constant, $\alpha_s(r) = \mathrm{const}/\ln{(r\Lambda_\mathrm{QCD}
/\hbar c)}$, where $r$ has dimension of length.

In the Standard Model (SM) there is another fundamental dimensional
parameter --- the Higgs vacuum expectation value (VEV), which
determines electroweak unification scale. Electron mass $m_e$ and
quark masses $m_q$ are proportional to the Higgs VEV. Consequently,
the dimensionless parameters $X_e=m_e/\Lambda_\mathrm{QCD}$ and
$X_q=m_q/\Lambda_\mathrm{QCD}$ link electroweak unification scale
with strong scale. For the light quarks $u$ and $d$, $X_q\ll 1$.
Because of that the proton mass $m_p$ is proportional to
$\Lambda_\mathrm{QCD}$ and $X_e\propto \mu$. Below we will use $\mu$
instead of $X_e$ because it is more directly linked to
experimentally measured atomic and molecular observables.

Below we will show that huge enhancement of the relative variation
happens in transitions between close atomic, molecular and nuclear
energy levels. Recently several new cases were found, where the
levels are very close and narrow. Large enhancement of the variation
effects is also possible in cold collisions of atoms and molecules
near Feshbach resonances.

We will start with general review of the present situation in the
search of the variation of $\alpha$ and $\mu$. After that we will
discuss in more detail the results, which follow from the
astrophysical observations of the optical and microwave spectra of
molecules. Finally, we will describe possible laboratory experiments
with molecules. This field is very new and there are no competitive
laboratory results on time-variation with molecules yet (see,
however, \sref{SF6}), but there are very promising proposals and
several groups already started experiments.

The analysis of the data from Big Bang nucleosynthesis
\cite{Dmitriev}, quasar absorption spectra, and Oklo natural nuclear
reactor give us the space-time variation of constants on the
Universe lifetime scale, i.e. on times from few billion to more than
ten billion years. Comparison of the frequencies of different atomic
and molecular transitions in laboratory experiments gives us the
present variation on the timescale from few months to few years.
There is no model independent connection between variations on such
different timescales. However, in order to compare the importance of
different results, we will often assume linear time dependence of
the constants. This way we can interpret all results in terms of
time derivatives of the fundamental constants. Within this
assumption, we can use quasar absorption spectra to obtain the best
current limit on the variation of the mass ratio $\mu$ and $X_e$
\cite{FK1}:
 \begin{align}\label{best_mu_dot}
 \dot{\mu}/\mu=\dot{X_e}/X_e=(1 \pm 3) \times
 10^{-16}\mathrm{~yr}^{-1}\,.
 \end{align}
We can also combine this result with the atomic clock results
\cite{clock_1} to get the best limit on variation of $\alpha$
\cite{tedesco,FS2007,Fla07}:
 \begin{align}\label{best_alpha_dot}
 \dot{\alpha}/\alpha=(-0.8 \pm 0.8) \times 10^{-16}\mathrm{~yr}^{-1}\,.
 \end{align}
Note that both limits \eqref{best_mu_dot} and \eqref{best_alpha_dot}
depend on the assumption of the linear time dependence of
fundamental constants.

The Oklo natural reactor gives the best limit on the variation of
$X_s=m_s/\Lambda_\mathrm{QCD}$ where  $m_s$ is the strange quark
mass \cite{Shl76,Oklo,FlambaumShuryak2002}:
 \begin{align}\label{best_X_s_dot}
 |\dot{X_s}/X_s| < 10^{-18}\mathrm{~yr}^{-1}\,.
 \end{align}
Note that the Oklo data can not give us any limit on the variation
of $\alpha$ since the effect of $\alpha$ there is much smaller than
the effect of $X_s$ and within the accuracy of the present theory
should be neglected \cite{FlambaumShuryak2002}.

In addition to the time-variation, one can also consider
spatial-variation of constants. Massive bodies (stars or galaxies)
can also affect physical constants.
%They have large scalar charge $S$ proportional to the number of
%particles which can produce a Coulomb-like scalar field $U=S/r$.
In other words the fundamental constants may depend on the
gravitational potential, e.g.
 \begin{align}\label{aG0}
 \delta \alpha/ \alpha = k_\alpha \delta (GM/ r c^2)\,,
 \end{align}
where $G$ is the gravitational constant and $r$ is the distance from
the mass $M$. The strongest limit on such variation:
 %\begin{subequations}\label{best_k_var}
 \begin{align}\label{aG}
 k_\alpha +0.17 k_\mu&= (-3.5\pm 6) \times 10^{-7}\,,
 %\\
 %k_\alpha +0.13 k_q&= (-1\pm 17) \times 10^{-7}\,,
 \end{align}
 %\end{subequations}
is obtained in Ref.~\cite{FS2007} from the measurements of the
dependence of atomic frequencies on the distance from the Sun due to
the ellipticity of the Earth's orbit \cite{clock_1,Ashby}
(parameters $k_\mu$ is defined by analogy with \Eref{aG0}). Below we
will also discuss some other results, including those, which
indicate nonzero variation of fundamental constants.

\section{Theoretical motivation}\label{motivation}

How changing physical constants and violation of local position
invariance may occur? Light scalar fields very naturally appear in
modern cosmological models, affecting  parameters of SM including
$\alpha$ and $\mu$ (for the whole list of SM parameters see
\cite{Wil07}). Cosmological variations of these scalar fields should
occur because of drastic changes of the composition of the Universe
during its evolution.

Theories unifying gravity and other interactions suggest the
possibility of spatial and temporal variation of physical
``constants'' in the Universe \cite{Marciano}. Moreover, there
exists a mechanism for making all coupling constants and masses of
elementary particles both space and time dependent, and influenced
by local environment (see review \cite{Uzan}). Variation of coupling
constants can be non-monotonic, such as damped oscillations, for
instance.

These variations are usually associated with the effect of massless
(or very light) scalar fields. One candidate is the dilaton: a
scalar which appears in string theories together with graviton, in a
massless multiplet of closed string excitations. Other scalars
naturally appear in cosmological models, in which our Universe is a
``brane'' floating in a space of larger dimensions. The scalars  are
simply brane coordinates in extra dimensions. However, the only
relevant scalar field recently discovered, the cosmological dark
energy, so far does not show visible variations. Observational
limits on the variations of physical constant given in \sref{intro}
are quite stringent, allowing only scalar couplings, which are tiny
in comparison with gravity.

A possible explanation was suggested by Damour \etal\
\cite{Damour1,Damour:1994zq} who pointed out that  cosmological
evolution of scalars naturally leads to their self-decoupling.
Damour and Polyakov have further suggested that variations should
happen when the scalars get excited by some physical change in the
Universe, such as phase transitions, or other drastic changes in the
equation of state of the Universe. They considered several of them,
but since the time of their paper a new fascinating transition has
been discovered: from matter dominated (decelerating) era to
dark-energy dominated (accelerating) era. It is a relatively recent
event, corresponding to cosmological redshift $z\approx 0.5$, or the
look-back time of approximately 5 billion years.

The time dependence of the perturbation related to this transition
can be calculated, and it turned out \cite{Barrow,Olive} that the
self-decoupling process is effective enough to explain why after
this transition the variation of constants is as small as observed
in laboratory experiments at the present time, while being at the
same time consistent with possible observations of the variations of
the electromagnetic fine-structure constant at $z\gtrsim 1$
\cite{Murphy,Ubach,LML07}.

\section{Dependence of atomic and molecular spectra on $\alpha$ and $\mu$}

Atomic and molecular spectra are most naturally described in atomic
units $(\hbar=m_e=e=1)$, where energy is measured in Hartrees (1
Hartree = $\tfrac{e^4 m_e}{\hbar^2}$ = 2~Ry =
219474.6313705(15)~cm$^{-1}$). In these units nonrelativistic
Schr\"{o}dinger equation for an atom with infinitely heavy pointlike
nucleus does not include any dimensional parameters. The dependence
of the spectrum on $\alpha$ appears only through relativistic
corrections, which describe fine-structure, Lamb shift, etc. The
dependence of atomic energies on $\mu$ is known as isotope effect
and is caused by finite nuclear mass and volume. There are even
smaller corrections to atomic energies, which depend on both
$\alpha$ and $\mu$ and are known as hyperfine structure.

One can argue that atomic energy unit itself depends on $\alpha$ as
it can be expressed as $\alpha^2m_e c^2$, where $m_e c^2$ is the
rest energy of the free electron. However, experimental search for
possible variation of fundamental constants consists in observing
time-variations of the ratios of different transition frequencies to
each other. In such ratios the dependence of the units on
fundamental constants cancels out. Below we will use atomic units
unless otherwise is explicitly stated.

Relativistic corrections to the binding energies of atomic valence
electrons are of the order of $\alpha^2 Z^2$, where $Z$ is atomic
number and become quite large for heavy elements. For our purposes,
it is convenient to present the dependence of atomic transition
frequencies on $\alpha^2$ in the form
 \begin{equation}\label{q-factor}
 \omega = \omega_0 + q  x,
 \end{equation}
where $x = (\frac{\alpha}{\alpha_0})^2 - 1 \approx \frac{2 \delta
\alpha}{\alpha}$ and $\omega_0 $ is a transition frequency for
$\alpha=\alpha_0$. Rough estimates of $q$-factors can be obtained
from simple one-particle models, but in order to obtain accurate
values one has to account for electronic correlations and perform
large-scale numerical calculations. Recently such calculations were
made for many atoms and ions
\cite{dzuba1999,Dy,q,nevsky,BEIK06,PKT07,DJ07,DF07b}.

Isotope effects in atoms are of the order of $\mu\sim 10^{-3}$ and
magnetic hyperfine structure roughly scales as $\alpha^2 \mu Z
g_\mathrm{nuc} \sim 10^{-7}Zg_\mathrm{nuc}$, where $g_\mathrm{nuc}$
is nuclear $g$-factor. One has to keep in mind that $g_\mathrm{nuc}$
also depends on $\mu$ and quark parameters $X_q$. This dependence
has to be considered, when we compare, for example, the frequency of
the hyperfine transition in $^{133}$Cs (Cs frequency standard)
\cite{tedesco}, or the hydrogen 21~cm hyperfine line
\cite{TWM05,TWM07} to various optical transitions \cite{tedesco}.

At present there are many very accurate experiments where different
optical and microwave atomic clocks are compared to each other
\cite{clock_1,prestage,Marion2003,Bize2005,Peik2004,Bize2003,
Fischer2004,Peik2005,Peik2006}. These experiments place strong
limits on the time-variation of different combinations of $\alpha$,
$\mu$, and $g_\mathrm{nuc}$. As we mentioned above, the limit on
$\alpha$-variation \eqref{best_alpha_dot} follows from the
experiment \cite{clock_1} and the limit \eqref{best_mu_dot} in the
assumption of linear time-dependence of all constants. A detailed
discussion of atomic experiments can be found in recent reviews
\cite{karshenboim,Lea07}.

On a cosmological timescale a comparison of the hyperfine transition
in atomic hydrogen with optical transitions in ions, was done in
Refs.~\cite{TWM05,TWM07}. This method allows one to study
time-variation of the parameter $F=\alpha^2g_p\mu$, where $g_p$ is
proton $g$-factor. Analysis of the absorbtion spectra of nine
quasars with redshifts $0.23 \le z \le 2.35$ gave
\begin{align}\label{x_var}
    \delta F/F &=(6.3\pm 9.9)\times 10^{-6},\\
\label{x_vara}
     \dot{F}/F &=(-6\pm 12)\times 10^{-16}~\mathrm{yr}^{-1},
\end{align}
which is consistent with zero variation of $\mu$ and $\alpha$.

Molecular spectroscopy opens additional possibilities to study
variation of fundamental constants. It is known that $\mu$ defines
the scales of electronic, vibrational, and rotational intervals in
molecular spectra, $E_\mathrm{el} : E_\mathrm{vib} : E_\mathrm{rot}
\sim 1 : \mu^{1/2} : \mu$. In addition to that, molecules also have
fine and hyperfine structure, $\Lambda$-doubling, hindered
rotations, etc. All these structures have different dependencies on
fundamental constants. Obviously, comparison of these structures to
each other allows the study of different combinations of fundamental
constants.

The sensitivity to temporal variation of fundamental constants may
be strongly enhanced in transitions between narrow close levels of
different nature. Huge enhancement of the relative variation
$\delta\omega/\omega$ can be obtained in transition between almost
degenerate levels in atoms
\cite{dzuba1999,Dy,nevsky,budker,budker1}, molecules
\cite{DeM04,mol,VKB04,FK1,FK2}, and nuclei \cite{th,th4}.

%\section[Feshbach resonances]{Enhancement of variation of
%fundamental constants in ultracold atom and molecule systems near
%Feshbach resonances}

An interesting case of enhancement of the effect of variation of
fundamental constants can be found in collisions of ultracold atoms
and molecules near Feshbach resonances \cite{chin}. The scattering
length $A$ near the resonance is extremely sensitive to the
$\mu$-variation:
 \begin{equation}\label{d_a_final}
 \frac{\delta A}{A}=K\frac{\delta\mu}{\mu}\,,
 \end{equation}
where the enhancement factor $K$ can be very large. For example, for
Cs-Cs collisions $K\sim 400$ \cite{chin}. Enhancement can be further
increased by adjusting the position of the resonance using external
fields. Near a narrow magnetic or optical Feshbach resonance the
enhancement factor $K$ may be increased by many orders of magnitude.

Calculation of the factor $K$ in Ref.~\cite{chin} is based on the
analytical formula for the scattering length derived in Ref.
\cite{Gribakin}. This formula is valid for an arbitrary interatomic
potential with power long-range tale ($1/r^6$ for neutral atoms),
i.e. this result includes all unharmonic corrections.

To the best of our knowledge, it is the only suggested experiment on
time-variation, where the observable is not frequency. Because of
that, we have to find another parameter $L$ of the dimension of
length to compare $A$ with. In Ref. \cite{chin} the scattering
length was defined in atomic units $(a_B)$. It is important,
however, that because of the large enhancement in \Eref{d_a_final},
the possible dependence of $L$ on $\mu$ becomes irrelevant. For
example, if we measure $A$ in conventional units, meters, which are
linked to Cs standard, then $\delta L/L =-\delta\mu/\mu$, and
 \begin{equation}\label{d_aL}
 \frac{\delta (A/L)}{(A/L)}=(K+1)\frac{\delta\mu}{\mu}\,.
 \end{equation}
As long as $K\gg 1$ the dependence of the used units on fundamental
constants can be neglected. Below we will discuss several other
experiments with huge enhancement factors, where this argument can
be also applied.

%-------------------------------------------------------------------------

\section{Astrophysical observations of the spectrum of \uppercase{H}$_2$}
\label{secH2}

{H}$_2$ is the most common molecule in the Universe and its UV
spectra have been used for the studies of the possible
$\mu$-variation for a long time. For a given electronic transition,
the frequency of each rovibrational line has different dependence on
$\mu$ \cite{Tho75,VL93}. Therefore, comparison of rovibrational
frequencies from astrophysics with laboratory observations can give
information on $\mu$.

In the adiabatic approximation, the rovibrational levels of the
electronic state $\Lambda$ with vibrational and rotational quantum
numbers $v$ and $J$ are given by the Dunham expression \cite{Dun32}:
 \begin{align}\label{Dunham1}
 E(v,J) &= \sum_{k,l\ge 0} Y_{k,l}
 \left(v+\tfrac12\right)^k \left[J(J+1)-\Lambda^2\right]^l,
 \end{align}
where each term depends on $\mu$ in a following way:
 \begin{align}\label{Dunham2}
 Y_{k,l} &\propto \mu^{l+k/2}.
 \end{align}
Because of the smallness of the parameter $\mu$, coefficients
$Y_{k,l}$ rapidly decrease with both $k$ and $l$, and for small $v$
and $J$, we have the usual vibrational ($k=1$) and rotational
($l=1$) terms. The zero term of this expansion ($k=l=0$) corresponds
to the electronic energy.

One can define the sensitivity coefficient $K_i$ for each
rovibrational transition $i$ of a given electronic band $e-g$
\cite{VL93}:
 \begin{align}
 K_i &\equiv\left(\frac{d\nu_i}{\nu_i}\right)
 \Big/\left(\frac{d\mu}{\mu}\right)
 \nonumber \\
 \label{Ki}
 &=\frac{\mu}{E_e-E_g}
 \left(\frac{dE_e}{d\mu}-\frac{dE_g}{d\mu}\right),
 \end{align}
where both energies are given by expansion \eqref{Dunham1}. The sign
of $K_i$ depends on the rovibrational energies of the excited ($e$)
and ground ($g$) states (in the absorbtion spectra of the quasars
only transitions from the ground electronic state are seen). The
electronic energy, presented by the term $Y_{0,0}$, dominates the
expansion and the coefficients $K_i$ are rather small. Typically
they are on the order $10^{-2}$, but can reach 0.05 for large values
of the quantum numbers $v$ and $J$.

The coefficients of expansion \eqref{Dunham1} can be found by
fitting experimental spectra. After that the sensitivity
coefficients $K_i$ are found from Eqs.~\eqref{Dunham2} and
\eqref{Ki}. Some rovibrational levels of different electronic
excited states appear to be very close. For such levels additional
non-adiabatic corrections can be included within the two-level
approximation \cite{HHS94}.

If there is $\mu$-variation $\Delta \mu$, this would lead to a
difference in observed redshifts $z_i$ for different lines:
 \begin{align}\label{dzeta}
 \zeta_i &\equiv
 \frac{z_i-z_\mathrm{q,abs}}{1+z_\mathrm{q,abs}}
 =-\frac{\Delta\mu}{\mu}K_i.
 \end{align}
By plotting the reduced redshifts $\zeta_i$ against the sensitivity
coefficients $K_i$, one can estimate $\Delta\mu/\mu$. The most
recent study \cite{Ubach} of the possible $\mu$-variation using
astrophysical data on H$_2$ was based on the observation of the two
quasar absorbtion systems with redshifts $z_\mathrm{q,abs}=3.02$ and
$2.59$. An analysis of the data on 76 lines from two UV bands of
H$_2$ gave the following result:
 \begin{align}\label{ubach_result1}
 \frac{\Delta\mu}{\mu}&=(-20\pm 6)\times 10^{-6}.
 \end{align}
This result indicates, at a $3.5\sigma$ confidence level, that $\mu$
has increased during the past 12 billion years. Assuming linear
time-dependence we can rewrite \Eref{ubach_result1} as
 \begin{align}\label{ubach_result2}
 \frac{\dot{\mu}}{\mu}&=(17\pm 5)\times 10^{-16}~\mathrm{yr}^{-1}.
 \end{align}
This has to be compared with the ammonia result \eqref{best_mu_dot},
which corresponds to a timescale about 6.5 billion years and is
discussed in more detail in \sref{secNH3}.

\section{Astrophysical observations of microwave molecular spectra}
\label{micro}

In the previous section we discussed astrophysical observations of
UV spectra of H$_2$. The corresponding absorbtion bands are very
strong and can be observed even for objects with very high
redshifts. On the other hand, as we have seen, the sensitivity
coefficients $K_i$ in \Eref{Ki} are rather small. This is because of
the relative smallness of rovibrational energy compared to the total
transition energy. Thus, it may be useful to study microwave spectra
of molecules, where the relative frequency variations due to varying
constants are larger.

\subsection{Rotational spectra}
\label{rot}

In 1996 Varshalovich and Potekhin \cite{VP96} compared redshifts for
microwave rotational transitions $(J=3 \rightarrow J=2)$ and $(J=2
\rightarrow J=1)$ in the CO molecule with redshifts of optical lines
of light atomic ions from the same astrophysical objects at
redshifts $z=2.286$ and $z=1.944$. As long as atomic frequencies are
independent on $\mu$ and rotational transition frequencies are
proportional to $\mu$, this comparison allowed to set the following
limits on variation of $\mu$:
\begin{subequations}\label{rotCO1}
\begin{align}
 \frac{\delta \mu}{\mu} &= (-0.6\pm 3.7)\times 10^{-4}
 &\mathrm{at}\,\, z=2.286\,,\\
 \frac{\delta \mu}{\mu} &= (-0.7\pm 1.0)\times 10^{-4}
 &\mathrm{at}\,\, z=1.944\,.
\end{align}
\end{subequations}

In the same paper \cite{VP96}, the authors compared the $(J=0
\rightarrow J=1)$ CO absorbtion line with the 21~cm hydrogen line
for an object with $z=0.2467$. They did not find a significant
difference in respective redshifts and interpreted this result as
yet another limit on variation of $\mu$. However, as we mentioned
above, the frequency of the hydrogen hyperfine line is proportional
to $\alpha^2\mu g_p$, and this result actually places limit on the
variation of the parameter $F=\alpha^2 g_p$~\cite{DWB98}. Recently a
similar analysis was performed by Murphy \etal\ \cite{MWF01d} using
more accurate data for the same object at $z=0.247$ and for a more
distant object at $z=0.6847$, and the following limits were
obtained:
\begin{subequations}\label{rotCO2}
\begin{align}
 \frac{\delta F}{F} &= (-2.0\pm 4.4)\times 10^{-6}
 &\mathrm{at}\,\, z=0.2467\,,\\
 \frac{\delta F}{F} &= (-1.6\pm 5.4)\times 10^{-6}
 &\mathrm{at}\,\, z=0.6847\,.
\end{align}
\end{subequations}
The object at $z=0.6847$ is associated with the gravitational lens
toward quasar B0218+357 and corresponds to the backward time $\sim
6.5$ Gyr. This object was also used by other authors, as will be
discussed in \sref{secOH} and \sref{secNH3}.

\subsection{The 18~\lowercase{cm} transitions in \uppercase{OH}}
\label{secOH}

Let us consider transitions between hyperfine substates of the
${}^2\Pi_{3/2}$ ground-state $\Lambda$-doublet in the OH molecule
\cite{Dar03,CK03,KCG04}. The $\Lambda$-doubling for ${}^2\Pi_{3/2}$
states appears in the third order in the Coriolis interaction and
the corresponding energy interval is inversely proportional to the
spin-orbit splitting between the ${}^2\Pi_{3/2}$ and ${}^2\Pi_{1/2}$
states, i.e., it scales as $\mu^3\alpha^{-2}$, while the hyperfine
structure intervals scale as $\alpha^2\mu g_\mathrm{nuc}$.
Therefore, the ratio of the hyperfine interval to the
$\Lambda$-doubling interval depends on the combination
$\tilde{F}=\alpha^4\mu^{-2} g_\mathrm{nuc}$. Higher-order
corrections modify this parameter to the form
$\tilde{F}=\alpha^{3.14}\mu^{-1.57} g_\mathrm{nuc}$ \cite{KCL05}.

The hyperfine-structure splitting for the OH molecule is
approximately 50~MHz and is much smaller than $\Lambda$-doubling
interval, which is about 1700~MHz. Because of that, it is actually
easier to compare the $\Lambda$-doubling transitions in OH to the
21~cm hydrogen line, or to rotational lines of the HCO$^+$ molecule
\cite{Dar03,CK03,KCG04,KCL05}.

The most stringent limit on the variation of $\tilde{F}$ was
obtained in Ref.~\cite{KCL05} from observations of the $z=0.6847$
gravitational lens:
 \begin{align}\label{OH1}
 \Delta \tilde{F}/\tilde{F} = \left(0.44 \pm0.36^\mathrm{stat}\pm
 1.0^\mathrm{syst}\right)\times 10^{-6}\,,
 \end{align}
where systematic error mostly accounts for the possible Doppler
noise, i.e. for the possible difference in the velocity
distributions of different molecules in a molecular cloud.

The laboratory frequencies of the OH $\Lambda$-doublet were recently
remeasured with higher precision using cold molecules produced by a
Stark decelerator \cite{HLS06}. That may become important for future
astrophysical measurements with higher accuracy.

\section{Limit on time-variation of $\mu$ from
inversion spectrum of ammonia} \label{secNH3}

Several years ago, van Veldhoven \etal\ suggested to use decelerated
molecular beam of ND$_3$ to search for the variation of $\mu$ in
laboratory experiments \cite{VKB04}. The ammonia molecule has a
pyramidal shape and the inversion frequency depends on the
exponentially small probability of tunneling of the three hydrogen
(or deuterium) atoms through the potential barrier \cite{TS55}.
Because of that, it is very sensitive to any changes of the
parameters of the system, particularly to the reduced mass for this
vibrational mode. The authors of Ref.~\cite{VKB04} found that for
ND$_3$ molecule, $\delta \omega/\omega=-5.6\, \delta \mu/\mu$.
Therefore, the inversion frequency of ND$_3$ is nearly an order of
magnitude more sensitive to $\mu$-variation than typical molecular
vibrational frequencies (note that Ref.~\cite{VKB04} contains a
misprint in the sign of the effect).

However, even such enhanced sensitivity is insufficient to make the
laboratory experiment on the time-variation of $\mu$ using
conventional molecular beams competitive. Stark-deceleration was
used in Ref.~\cite{VKB04} to slow down the beam to 52 m/s. Still, a
much slower beam, or a fountain is necessary to increase the
sensitivity by several orders of magnitude before a competitive
experiment can be performed. The work in this direction is in
progress \cite{Bet07}.

On the other hand, an only slightly smaller enhancement also exists
for the inversion spectrum of NH$_3$, which is often seen in
astrophysics, even for high $z$ objects. This fact was used in
\cite{FK1} to place the limit \eqref{best_mu_dot}, which we will now
discuss in some detail. The inversion vibrational mode of ammonia is
described by a double-well potential with the first two vibrational
levels lying below the barrier. Because of the tunneling, these two
levels are split in inversion doublets. The lower doublet
corresponds to the wavelength $\lambda\approx 1.25$~cm and is used
in ammonia masers. Molecular rotation leads to the centrifugal
distortion of the potential curve. Because of that, the inversion
splitting depends on the rotational angular momentum $J$ and its
projection $K$ on the molecular symmetry axis:
 \begin{align}\label{w_inv}
 \omega_\mathrm{inv}(J,K) = \omega^0_\mathrm{inv}
 - c_1
 \left[J(J+1)-K^2\right] + c_2 K^2 + \cdots \,,
 \end{align}
where we omitted terms with higher powers of $J$ and $K$.
Numerically, $\omega^0_\mathrm{inv}\approx 23.787$~GHz, $c_1\approx
151.3$~MHz, and $c_2\approx 59.7$~MHz.

In addition to the rotational structure \eqref{w_inv} the inversion
spectrum includes much smaller hyperfine structure. For the main
nitrogen isotope $^{14}$N, the hyperfine structure is dominated by
the electric quadrupole interaction ($\sim 1$~MHz)~\cite{HT83}.
Because of the dipole selection rule $\Delta K=0$ the levels with
$J=K$ are metastable and in laboratory beam experiments the width of
the corresponding inversion lines is usually determined by
collisional broadening. In astrophysics the lines with $J=K$ are
also narrower and stronger than others, but the hyperfine structure
for spectra with high redshifts is still unresolved.

For our purposes it is important to know how the parameters in
\Eref{w_inv} depend on fundamental constants. The molecular
electrostatic potential in atomic units does not depend on
fundamental constants (here we neglect small relativistic
corrections which give a weak $\alpha$ dependence). Therefore, the
inversion frequency $\omega^0_\mathrm{inv}$ and constants $c_{1,2}$
are functions of $\mu$ only. Note that the coefficients $c_i$ depend
on $\mu$ through the reduced mass of the inversion mode and because
they are inversely proportional to the molecular moments of inertia.
This implies a different scaling of $\omega^0_\mathrm{inv}$ and
$c_i$ with $\mu$.

The inversion spectrum \eqref{w_inv} can be approximately described
by the following Hamiltonian:
\begin{align}\label{H_inv1}
 H_\mathrm{inv} &= -\tfrac{1}{2M_1} \partial^2_x+U(x)\\
 &+\tfrac{1}{I_1(x)}\left[J(J+1)-K^2\right]
 +\tfrac{1}{I_2(x)}K^2,
\nonumber
\end{align}
where $x$ is the distance from N to the H-plane, $I_1$, $I_2$ are
moments of inertia perpendicular and parallel to the molecular axis,
correspondingly, and $M_1$ is the reduced mass for the inversion
mode. If we assume that the length $d$ of the N---H bond does not
change during inversion, then $M_1=2.54\,m_p$ and
\begin{align}\label{moment_I1}
 I_1(x) &\approx \tfrac{3}{2}m_p d^2 \left[1+0.2(x/d)^2\right], \\
 \label{moment_I2}
 I_2(x) &\approx 3m_p d^2 \left[1-(x/d)^2\right].
\end{align}
The dependence of $I_{1,2}$ on $x$ generates correction to the
potential energy of the form $C(J,K)\,x^2\mu$. This changes the
vibrational frequency and the effective height of the potential
barrier, therefore changing the inversion frequency
$\omega_\mathrm{inv}$ given by \Eref{w_inv}.

Following Ref.~\cite{SI62} we can write the potential $U(x)$ in
\Eref{H_inv1} in the following form:
\begin{align}\label{H_inv2}
 U(x) &= \tfrac{1}{2}k x^2 +b \exp\left(-c x^2\right).
\end{align}
Fitting vibrational frequencies for NH$_3$ and ND$_3$ gives
$k\approx 0.7598$~a.u., $b\approx 0.05684$~a.u., and $c\approx
1.3696$~a.u. Numerical integration of the Schr\"{o}dinger equation
with the potential \eqref{H_inv2} for different values of $\mu$
gives the following result:
\begin{align}
 \label{dw_inv6}
 \frac{\delta\omega_\mathrm{inv}^0}{\omega_\mathrm{inv}^0}
 &\approx 4.46\, \frac{\delta\mu}{\mu}\,.
\end{align}
It is instructive to reproduce this result from an analytical
calculation. In the WKB approximation the inversion frequency is
estimated as \cite{LL77}:
\begin{subequations}
\begin{align}
 \label{w_inv1}
 \omega_\mathrm{inv}^0
 &= \frac{\omega_\mathrm{vib}}{\pi}\exp\left(-S\right) \\
 \label{w_inv2}
 &= \frac{\omega_\mathrm{vib}}{\pi}
 \exp\!\left(-\frac{1}{\hbar}\int_{-a}^a\! \sqrt{2M_1(U(x)-E)}\,
 \mathrm{d} x\right),
\end{align}
\end{subequations}
where $\omega_\mathrm{vib}$ is the vibrational frequency of the
inversion mode, $S$ is the action in units of $\hbar$, $x=\pm a$ are
classical turning points for the energy $E$. For the lowest
vibrational state $E=U_\mathrm{min}+\tfrac12 \omega_\mathrm{vib}$.
Using the experimental values $\omega_\mathrm{vib}=950$~cm$^{-1}$
and $\omega_\mathrm{inv}=0.8$~cm$^{-1}$, we get $S\!\approx\!
5.9\,$.

Expression \eqref{w_inv2} allows one to calculate the dependence of
$\omega_\mathrm{inv}^0$ on the mass ratio $\mu$. Let us present $S$
in the following form: $S = A\mu^{-1/2}\int_{-a}^{a}
\sqrt{U(x)-E}\,\mathrm{d}x$, where $A$ is a numerical constant and
the square root depends on $\mu$ via $E$:
\begin{subequations}
\begin{align}
 \label{dw_inv1}
 \frac{\mathrm{d}\omega_\mathrm{inv}^0}{\mathrm{d}\mu}
 &= \omega_\mathrm{inv}^0
 \left(\frac{1}{2\mu}-
 \frac{\mathrm{d}S}{\mathrm{d}\mu} \right)\\
 \label{dw_inv2}
 &= \omega_\mathrm{inv}^0
 \left(\frac{1}{2\mu}-
 \frac{\partial S}{\partial\mu}
 -\frac{\partial S}{\partial E}\frac{\partial E}{\partial\mu}
 \right).
\end{align}
\end{subequations}

It is easy to see that ${\partial S}/{\partial\mu}=-S/2\mu$. The
value of the third term in \Eref{dw_inv2} depends on the form of the
potential barrier:
\begin{align}
 \label{dw_inv3}
 \frac{\partial S}{\partial E}
 &=-\frac{q}{4}\frac{S}{U_\mathrm{max}-E},
\end{align}
where for a square barrier $q=1$, and for a triangular barrier
$q=3$. For a more realistic barrier shape, $q\approx 2$. Using the
parametrization \eqref{H_inv2} to determine $U_\mathrm{max}$ we get:
\begin{align}
 \label{dw_inv4}
 \!\frac{\delta\omega_\mathrm{inv}^0}{\omega_\mathrm{inv}^0}
 &\approx \frac{\delta\mu}{2\mu}
 \left(1+S
 +\frac{S}{2}\frac{\omega_\mathrm{vib}}{U_\mathrm{max}-E}
 \right)
 = 4.4\, \frac{\delta\mu}{\mu},
\end{align}
which is close to the numerical result \eqref{dw_inv6}.

We see that the inversion frequency of NH$_3$ is an order of
magnitude  more sensitive to the change of $\mu$ than typical
vibrational frequencies. The reason for this is clear from
\Eref{dw_inv4}: it is the large value of the action $S$ for the
tunneling process.

Using Eqs.~\eqref{H_inv1} --~\eqref{moment_I2} one can also find the
dependence on $\mu$ of the constants $c_{1,2}$ in \Eref{w_inv}
\cite{FK1}:
 \begin{align}
 \label{dw_rot5}
 \frac{\delta c_{1,2}}{c_{1,2}}
 &= 5.1\frac{\delta\mu}{\mu}\,.
 \end{align}

It is clear that the above consideration is directly applicable to
ND$_3$, where the inversion frequency is 15 times smaller and
\Eref{w_inv2} gives $S\approx 8.4$. According to \Eref{dw_inv4},
this leads to a somewhat higher sensitivity of the inversion
frequency to $\mu$ in agreement with Ref.~\cite{VKB04}:
\begin{align}
 \label{dw_nd3}
 \mathrm{ND_3:}
 \left\{
 \begin{array}{rcl}
   \frac{\delta\omega_\mathrm{inv}}{\omega_\mathrm{inv}}
   &\approx &5.7\, \frac{\delta\mu}{\mu}\,,\\
\\
   \frac{\delta c_2}{c_2}
   &\approx &6.2\,\frac{\delta\mu}{\mu}\,.
 \end{array}
 \right.
\end{align}

We see from Eqs.~\eqref{dw_inv6} and \eqref{dw_rot5} that the
inversion frequency $\omega_\mathrm{inv}^0$ and the rotational
intervals
$\omega_\mathrm{inv}(J_1,K_1)-\omega_\mathrm{inv}(J_2,K_2)$ have
different dependencies on $\mu$. In principle, this allows one to
study time-variation of $\mu$ by comparing different intervals in
the inversion spectrum of ammonia. For example, if we compare the
rotational interval to the inversion frequency, then
Eqs.~\eqref{dw_inv6} and~\eqref{dw_rot5} give:
\begin{align}
 \label{red1}
 \frac{\delta\{[\omega_\mathrm{inv}(J_1,K_1)-\omega_\mathrm{inv}(J_2,K_2)]
 /\omega^0_\mathrm{inv}\}}
 {[\omega_\mathrm{inv}(J_1,K_1)-\omega_\mathrm{inv}(J_2,K_2)]/\omega^0_\mathrm{inv}}
 &= 0.6 \frac{\delta\mu}{\mu}\,.
\end{align}
The relative effects are substantially larger if we compare the
inversion transitions with the  transitions between the quadrupole
and magnetic hyperfine components. However, in practice, this method
will not work because of the smallness of the hyperfine structure
compared to typical linewidths in astrophysics.

Again, as in the case of $\Lambda$-doubling in OH molecule, it is
more promising to compare the inversion spectrum of NH$_3$ with
rotational spectra of other molecules, where
\begin{align}
 \label{red2}
 \frac{\delta\omega_\mathrm{rot}}{\omega_\mathrm{rot}}
 &= \frac{\delta\mu}{\mu}\,.
\end{align}
In astrophysics any frequency shift is related to a corresponding
apparent redshift:
\begin{align}
 \label{red3}
 \frac{\delta\omega}{\omega}
 &= - \frac{\delta z}{1+z}\,.
\end{align}
According to Eqs.~\eqref{dw_inv6} and~\eqref{red2}, for a given
astrophysical object with $z=z_0$ variation of $\mu$ leads to a
change of the apparent redshifts of all rotational lines $\delta
z_\mathrm{rot}=-(1+z_0)\,{\delta\mu}/{\mu}$ and the corresponding
shifts of all inversion lines of ammonia are: $\delta
z_\mathrm{inv}= -4.46\,(1+z_0)\,{\delta\mu}/{\mu}$. Therefore,
comparing the apparent redshift $z_\mathrm{inv}$ for NH$_3$ with the
apparent redshifts $z_\mathrm{rot}$ for rotational lines we can find
${\delta\mu}/{\mu}$:
\begin{align}
 \label{red4}
 \frac{\delta\mu}{\mu}
 &= 0.289\, \frac{z_\mathrm{rot}-z_\mathrm{inv}}{1+z_0}\,.
\end{align}

High-precision data on the redshifts of NH$_3$ inversion lines exist
for the already mentioned object B0218+357 at $z\approx 0.6847$
\cite{HJK05}. Comparing them with the redshifts of rotational lines
of CO, HCO$^+$, and HCN molecules from Ref.~\cite{CW97} one can get
the following conservative limit from \Eref{red4}:
\begin{equation} \label{nh3final}
 \frac{\delta\mu}{\mu}=(-0.6 \pm 1.9)\times 10^{-6}.
\end{equation}
Taking into account that the redshift $z\approx 0.68$ for the object
B0218+357 corresponds to the look-back time of about 6.5 Gyr, this
limit translates into the most stringent present limit
\eqref{best_mu_dot} for the variation rate $\dot\mu/\mu$.

\section{Experiment with \uppercase{SF}$_6$}\label{SF6}

Now we switch to laboratory molecular experiments on time-variation.
We start with the recent experiment on two-photon vibrational
transition $(v=0,J=4) \rightarrow (v=2,J=3)$ in SF$_6$ \cite{Cha07}.
This is a Ramsey-type experiment with a supersonic beam of SF$_6$
molecules. The beam velocity $u=400$~m/s and the length of the
interaction region $D=1$~m corresponds to the linewidth of
$u/2D=200$~Hz.

A CO$_2$ laser was used to drive the two-photon transition and its
frequency was controlled by a Cs standard \cite{AGG05}. This means,
that the vibrational frequency $\omega_\mathrm{vib}$ in SF$_6$ was
compared with the hyperfine transition frequency
$\omega_\mathrm{hfs}$ in Cs. Therefore, the experiment was sensitive
to the combination of fundamental constants $F = g_\mathrm{nuc}
\mu^{-1/2} \alpha^{2.83}$. Measurements continued for 18 months, and
the following result was obtained:
\begin{align}\label{SF6a}
 \dot{F}/F
 &= (1.4 \pm 3.2)\times 10^{-14}\, \mathrm{yr}^{-1}\,.
\end{align}

This limit is weaker, than the most stringent limits obtained with
atomic clocks. On the other hand, it constrains a different
combination of fundamental parameters. Most importantly, in atomic
experiments the parameters $g_n$ and $\mu$ always go as a product
$g_n\mu$, while here we have combination $g_n \mu^{-1/2}$. That
allows to combine atomic results \cite{Peik2004,Fischer2004,clock_1}
with limit \eqref{SF6a} to obtain the best laboratory limit on
$\mu$-variation:
\begin{align}\label{SF6b}
 \dot{\mu}/\mu
 &= (3.4 \pm 6.5)\times 10^{-14}\, \mathrm{yr}^{-1}\,.
\end{align}
This limit is significantly weaker than astrophysical limit
\eqref{best_mu_dot}, but there are good chances that it will be soon
significantly improved.

\section{Close narrow levels in diatomic molecules}\label{diatomics}

In this section we focus on very close narrow levels of different
nature in diatomic molecules. Such levels may occur due to
cancelation between either hyperfine and rotational structures
\cite{mol}, or between the fine and vibrational structures of the
electronic ground state \cite{FK2}. The intervals between the levels
correspond to microwave frequency range convenient for experiments
and the level widths are very small, typically $\sim 10^{-2}$~Hz.
The enhancement of the relative variation $K$ can exceed $10^5$.

\subsection{Molecules with cancelation between hyperfine
structure and rotational intervals} \label{hfs-rot}

Consider diatomic molecules with the unpaired electron and the
$^2\Sigma$ ground state. Examples of such molecules include LaS,
LaO, LuS, LuO, and YbF \cite{HH79}. The hyperfine interval
$\Delta_\mathrm{hfs}$ is proportional to $\alpha^2 Z
F_\mathrm{rel}(\alpha Z) \mu g_\mathrm{nuc}$, where $F_\mathrm{rel}$
is an additional relativistic (Casimir) factor \cite{Sob79}. The
rotational interval $\Delta_\mathrm{rot} \propto \mu$ is roughly
independent on $\alpha$. If we find a molecule with
$\Delta_\mathrm{hfs} \approx \Delta_\mathrm{rot}$, the splitting
$\omega$ between hyperfine and rotational levels will depend on the
following combination
 \begin{align}
 \label{hfs-rot1}
 \omega \propto \mu \left[\alpha^2 F_\mathrm{rel}(\alpha Z)\, g_\mathrm{nuc}
 - \mathrm{const}\right]\, .
 \end{align}
Relative variation is then given by
\begin{align}
\label{hfs-rot2}
 \frac{\delta\omega}{\omega}
 \approx \frac{\Delta_\mathrm{hfs}}{\omega}
 \left[\left(2+K\right)\frac{\delta\alpha}{\alpha} + \frac{\delta
 g_\mathrm{nuc}}{g_\mathrm{nuc}}\right]+\frac{\delta\mu}{\mu}\,,
\end{align}
where the factor $K$ comes from variation of $F_\mathrm{rel}(\alpha
Z)$, and for $Z \sim 50$, $K\approx 1$. As long as
$\Delta_\mathrm{hfs}/\omega\gg 1$, we can neglect the last term in
\Eref{hfs-rot2}.

The data on hyperfine structure of diatomics are sparse and usually
not very accurate. This hampers the search for molecules with strong
cancelation of the types, discussed here. Using data from
\cite{HH79} one can find that $\omega = (0.002\pm 0.01)$~cm$^{-1}$
for ${}^{139}$La${}^{32}$S \cite{mol}. Note that for $\omega =
0.002$~cm$^{-1}$ the relative frequency shift is:
\begin{align}
\label{hfs-rot3}
 \frac{\delta\omega}{\omega}
 \approx 600\,\frac{\delta\alpha}{\alpha}\,.
\end{align}
With new data on molecular hyperfine constants appearing regularly,
it is likely that other molecular candidates for such experiments
will appear soon.

% based on file 9b.tex (version 30/08/07 for arXiv) (includes more refs to e-prints)

\subsection{Molecules with cancelation between fine-structure
and vibrational intervals} \label{fs-vib}

The fine-structure interval $\omega_f$ rapidly grows with the
nuclear charge Z:
\begin{align}
\label{of}
 \omega_f \sim Z^2 \alpha^2\, ,
\end{align}
On the contrary, the vibration energy quantum decreases with the
atomic mass:
\begin{align}
 \label{ov}
\omega_\mathrm{vib} \sim M_r^{-1/2} \mu^{1/2}\, ,
\end{align}
where the reduced mass for the molecular vibration is $M_r m_p$.
Therefore, we obtain an equation $Z=Z(M_r,v)$ for the lines on the
plane $Z,M_r$, where we can expect approximate cancelation between
the fine-structure and vibrational intervals:
\begin{align}
 \label{o}
 \omega=\omega_f - v\,  \omega_\mathrm{vib} \approx 0 \,,
 \quad v=1,2,...
\end{align}
Using Eqs.~(\ref{of}--\ref{o}) it is easy to find the dependence of
the transition frequency on the fundamental constants:
\begin{align}
 \label{do}
 \frac{\delta\omega}{\omega}=
 \frac{1}{\omega}\left(2 \omega_f \frac{\delta\alpha}{\alpha}+
 \frac{v}{2} \omega_\mathrm{vib} \frac{\delta\mu}{\mu}\right)
 &\approx K \left(2 \frac{\delta\alpha}{\alpha}+
\frac{1}{2} \frac{\delta\mu}{\mu}\right),
\end{align}
where the enhancement factor $K= \frac{\omega_f}{\omega}$
%>mgk11/07 and 29/08>
determines the relative frequency shift for the given change of
fundamental constants. Large values of the factor $K$ hint at
potentially favorable cases for performing an experiment because it
is usually preferable to have larger relative shifts. However, there
is no strict rule that larger $K$ is always better. In some cases,
such as very close levels, this factor may become irrelevant
\cite{budker}. Thus, it is also important to consider the absolute
values of the shifts and compare them to the linewidths of the
corresponding transitions.

Because the number of molecules is finite we can not have $\omega=0$
exactly. However, for many molecules we do have $\omega/\omega_f \ll
1$ and $|K| \gg 1$. Moreover, an additional ``fine tuning'' may be
achieved by selection of isotopes and rotational, $\Omega$-doublet,
and hyperfine components. Therefore, we have two large manifolds,
the first one is built on the electron fine-structure excited state,
and the second one is built on the vibrational excited state. If
these manifolds overlap, one may select two or more transitions with
different signs of $\omega$. In this case expected sign of the
$|\omega|$-variation must be different (since the variation $\delta
\omega$ has the same sign) and one can eliminate some systematic
effects. Such control of systematic effects was used in Refs.
\cite{budker,budker1,Fer07} for transitions between close levels in
two dysprosium isotopes. The sign of energy difference between two
levels belonging to different electron configurations was opposite
for the $^{163}$Dy and $^{162}$Dy transitions used in that work.

\begin{table}[tbh]
  \caption{Diatomic molecules with quasidegeneracy between the
  ground-state vibrational and fine-structure excitations. All frequencies are
  in cm$^{-1}$. The data are taken from Ref.~\cite{HH79}. Enhancement
  factor $K$ is estimated using \Eref{do}.}
  \label{tab1}
  \begin{tabular}{llddr}
  \hline\hline
    Molecule    &Electronic states   &\multicolumn{1}{c}{$\omega_f$}
                                                &\multicolumn{1}{c}{$\omega_\mathrm{vib}$}
                                                          &\multicolumn{1}{c}{$K$} \\
  \hline
    Cl$_2^+$    &$^2\Pi_{3/2,1/2}   $&   645    &   645.6 & 1600 \\
    CuS         &$^2\Pi             $&   433.4  &   415   &   24 \\
    IrC         &$^2\Delta_{5/2,3/2}$&  3200    &  1060   &  160 \\
    SiBr        &$^2\Pi_{1/2,3/2}   $&   423.1  &   424.3 &  350 \\
  \hline\hline
  \end{tabular}
\end{table}

In \tref{tab1} we present the list of molecules from
Ref.~\cite{HH79} where the ground state is split in two
fine-structure levels and \Eref{o} is approximately fulfilled. The
molecules Cl$_2^+$ and SiBr are particularly interesting. For both
of them the frequency $\omega$ defined by \Eref{o} is of the order
of 1~cm$^{-1}$ and comparable to the rotational constant $B$. This
means that $\omega$ can be reduced further by proper choice of
isotopes, rotational quantum number $J$ and hyperfine components.
New dedicated measurements are needed to determine exact values of
the transition frequencies and to find the best transitions.
However, it is easy to find the necessary accuracy of the
frequency-shift measurements. According to \Eref{do}, the expected
frequency shift is
\begin{align}
\label{do1}
 \delta\omega=2 \omega_f \left(\frac{\delta\alpha}{\alpha}+
 \frac{1}{4}\frac{\delta\mu}{\mu}\right)\,.
\end{align}
Assuming $\delta \alpha / \alpha \sim 10^{-15}$ and $\omega_f\sim
500$~cm$^{-1}$, we obtain $\delta\omega \sim 10^{-12}$ cm$^{-1}\sim
3 \times 10^{-2}$ Hz. In order to obtain similar sensitivity
comparing hyperfine transition frequencies for Cs and  Rb one has to
measure the shift $\sim 10^{-5}$ Hz.

\subsection{Molecular ion {H\lowercase{f}F}$^+$}

The list of molecules in \tref{tab1} is incomplete because of the
lack of data in Ref.~\cite{HH79}. Let us briefly discuss one
interesting case, which has been brought to attention quite
recently. The HfF$^+$ ion and other similar ions are being
considered by E.~Cornell's group at JILA for an experiment to search
for the electric dipole moment (EDM) of the electron
\cite{SC04,MBD06}. In this experiment, the ions are to be trapped in
a quadrupole RF trap to achieve long coherence times. A similar
experimental setup can be used to study possible time-variation of
fundamental constants. A recent calculation by \citet{PMI06}
suggests that the ground state of this ion is $^1\Sigma^+$ and the
first excited state $^3\Delta_1$ lies only 1633~cm$^{-1}$ higher.
The calculated vibrational frequencies for these two states are 790
and 746~cm$^{-1}$, respectively. For these parameters the
vibrational level $v=3$ of the ground state is only 10~cm$^{-1}$
away from the $v=1$ level of the $^3\Delta_1$ state. Thus, instead
of \Eref{o} we now have:
\begin{align}
 \label{hff1}
 \omega=\omega_\mathrm{el} + \tfrac32 \omega_\mathrm{vib}^{(1)}
 - \tfrac72\omega_\mathrm{vib}^{(0)}\approx 0\,,
\end{align}
where superscripts 0 and 1 correspond to the ground and excited
electronic states. The electronic transition with frequency
$\omega_\mathrm{el}$ is not a fine-structure transition, and
\Eref{of} is not applicable. Instead, by analogy with
\Eref{q-factor} we can write:
\begin{align}
 \label{hff2}
 \omega_\mathrm{el}=\omega_\mathrm{el,0} + q x\,,
 \quad x=\alpha^2/\alpha_0^2-1\,.
\end{align}

In order to calculate the $q$-factor for HfF$^+$ ion one needs to
perform relativistic molecular calculations for several values of
$\alpha$, which has not been done yet. However, it is possible to
make an order of magnitude estimate using atomic calculation for the
Yb$^+$ ion \cite{Dy}. According to Ref.~\cite{PMI06}, the
$^1\Sigma_1^+$~--~$^3\Delta_1$ transition, to a first approximation,
corresponds to the $6s$~--~$5d$ transition in the hafnium ion. It is
well known that valence $s$- and $d$-orbitals of heavy atoms have
very different dependence on $\alpha$: while the binding energy of
$s$-electrons grows with $\alpha$, the binding energy of
$d$-electrons decreases \cite{dzuba1999,q,Dy,nevsky}. For the same
transition in the Yb$^+$ ion the Ref.~\cite{Dy} gives
$q_{sd}=10000$~cm$^{-1}$. Using this value as an estimate, we can
write by analogy with \Eref{do}:
\begin{align}
 \label{hff3}
 \frac{\delta\omega}{\omega}
 &\approx
 \left(\frac{2q}{\omega} \frac{\delta\alpha}{\alpha}+
 \frac{\omega_\mathrm{el}}{2\omega} \frac{\delta\mu}{\mu}\right)
 \approx \left(2000 \frac{\delta\alpha}{\alpha}+
 80 \frac{\delta\mu}{\mu}\right),
\\
 \label{hff4}
 \delta\omega
 &\approx 20000~\mathrm{cm}^{-1}(\delta\alpha/\alpha+0.04 \delta\mu/\mu)\,.
\end{align}
Assuming $\delta \alpha / \alpha \sim 10^{-15}$ we obtain
$\delta\omega \sim$ 0.6 Hz.

\subsection{Estimate of the natural widths of the quasidegenerate
states}\label{sec_lw}

%>mgk11/07>
As we mentioned above, it is important to compare frequency shifts
caused by time-variation of constants to the linewidths of
corresponding transitions.
%<mgk11/07<
First let us estimate the natural width $\Gamma_v$ of the
vibrational level $v$:
\begin{align}
 \label{lw1}
 \Gamma_v &= \frac{4\omega_\mathrm{vib}^3}{3\hbar c^3}|\langle
 v|\hat{D}|v-1\rangle|^2\,.
\end{align}
To estimate the dipole matrix element we can write:
\begin{align}
 \label{lw2}
 \hat{D}&=\left.\frac{\partial D(R)}{\partial
 R}\right|_{R=R_0}(R-R_0)
 \sim \frac{D_0}{R_0}(R-R_0)
 \,,
\end{align}
where $D_0$ is the dipole moment of the molecule for equilibrium
internuclear distance $R_0$. Using the standard expression for the
harmonic oscillator, $\langle v|x|v-1\rangle=\left(\hbar
v/2m\omega\right)^{1/2}$, we get:
\begin{align}
 \label{lw3}
 \Gamma_v &= \frac{2\omega_\mathrm{vib}^2 D_0^2 v}{3c^3M_r m_p R_0^2}\,.
\end{align}
For the homonuclear molecule Cl$_2^+$ $D_0=0$ and expression
\eqref{lw3} turns to zero. For the SiBr molecule, \Eref{lw3} gives
$\Gamma_1\sim 10^{-2}$~Hz, where we assume $D_0^2/R_0^2\sim
0.1\,e^2$.

Now let us estimate the width $\Gamma_f$ of the upper state of the
fine-structure doublet $^2\Pi_{1/2,3/2}$. By analogy with \Eref{lw1}
we can write:
\begin{align}
 \label{lw4}
 \Gamma_f &= \frac{4\omega_f^3}{3\hbar c^3}
 \left|\left\langle {}^2\Pi_{3/2}|D_1| {}^2\Pi_{1/2}\right\rangle\right|^2\,.
\end{align}
The dipole matrix element in this expression is written in the
molecular frame and we have summed over final rotational states.
This matrix element corresponds to a spin-flip and turns to zero in
the non-relativistic approximation. Spin-orbit interaction mixes the
${}^2\Pi_{1/2}$ and ${}^2\Sigma_{1/2}$ states:
\begin{align}
 \label{lw5}
  \left|{}^2\Pi_{1/2}\right\rangle &\rightarrow
  \left|{}^2\Pi_{1/2}\right\rangle
  + \xi\left|{}^2\Sigma_{1/2}\right\rangle
  \,,
\end{align}
and the matrix element in \Eref{lw4} becomes \cite{KFD87}:
\begin{align}
 \label{lw6}
 \left\langle {}^2\Pi_{3/2}|D_1| {}^2\Pi_{1/2}\right\rangle
 &\approx \xi \left\langle \Pi|D_1| \Sigma\right\rangle
 \sim \frac{\alpha^2 Z^2}{10(E_\Pi-E_\Sigma)}
 \,,
\end{align}
where $E_\Sigma$ is the energy of the lowest $\Sigma$-state.
Substituting \Eref{lw6} into \Eref{lw4} and using energies from
Ref.~\cite{HH79}, we get the following estimate for the molecules
Cl$_2^+$ and SiBr:
\begin{align}
 \label{lw7}
 \Gamma_f &\sim 10^{-2}~\mathrm{Hz}\,.
 %\left\{
 %\begin{array}{ll}
 %10^{-2}~\mathrm{Hz,}&\mathrm{Cl}_2^+\,, \\
 %10^{-1}~\mathrm{Hz,}&\mathrm{SiBr}\,.
 %\end{array}\right.
\end{align}
Here we took into account that the unpaired electron in SiBr
molecule is predominantly located at Si (Z=14) rather then at Br
(Z=35). Because of this, the fine-structure splitting in SiBr is
smaller than that of Cl$_2^+$, where $Z=17$ (see Table~\ref{tab1}).

%>mgk11/07>
We conclude that natural widths of the molecular levels considered
here are of the order of $10^{-2}$~Hz. This can be compared, for
example, to the natural width 12~Hz of the level $^2D_{5/2}$ of
Hg$^+$ ion, which was used in atomic experiment~\cite{clock_1}. For
such narrow levels the lifetime may depend on the interaction with
the black body radiation \cite{VD07}. According to this reference,
the lifetimes of the ro-vibrational levels of polar molecules at
room temperature vary from 1 s to 100 s.
%<mgk11/07<

%%%%%%%%%%%%%%%%%%%%%%%%%%%
%%%%%%%%%%%%%%%%%%%%%%%%%%%

\section{Experiments with C\lowercase{s}$_2$ and S\lowercase{r}$_2$}\label{Cs2}

In this section we discuss two recently proposed experiments with
cold diatomic molecules. The first one with Cs$_2$ molecule was
proposed at Yale \cite{DeM04,DSS07} and the second experiment with
Sr$_2$ molecule is in preparation at JILA \cite{ZKY07}.

\begin{figure*}[htb]
 \includegraphics[scale=1.0]{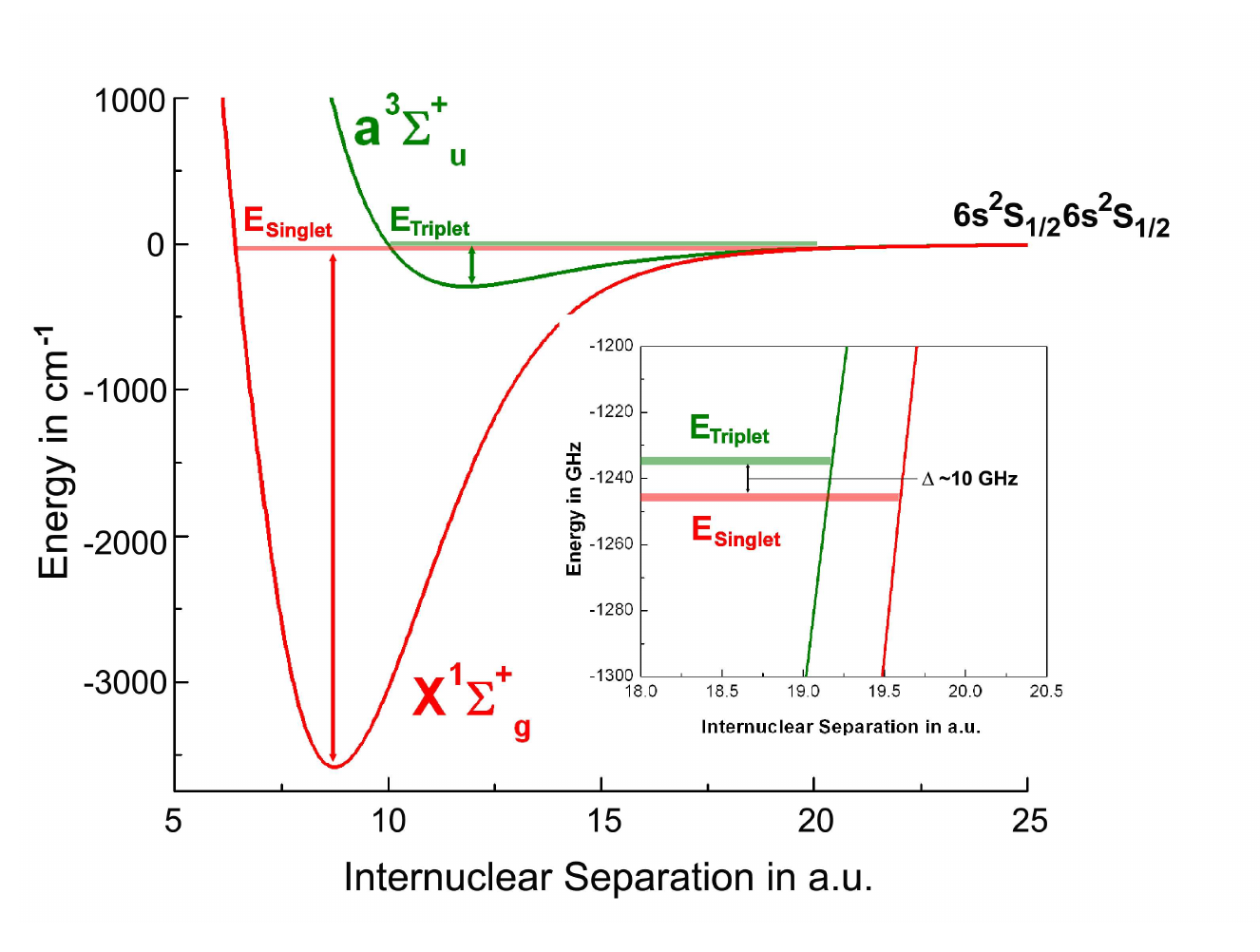}
 \caption{Levels $^3\Sigma_u^+$ and $^1\Sigma_g^+$ in Cs$_2$ molecule (figure
 from Ref.~\cite{Sai05}).} \label{fig_Cs2}
\end{figure*}

The Yale experiment is based on the idea \cite{DeM04} to match an
electronic energy with a large number of vibrational quanta. The
difference with Eqs.~(\ref{of}~--~\ref{o}) is that here electronic
transition is between the ground state $^1\Sigma_g^+$ and the
$^3\Sigma_u^+$ state and, to a first approximation, its frequency is
independent of $\alpha$. The energy of this transition is about
3300~cm$^{-1}$ and the number of vibrational quanta needed to match
this interval is on the order 100 (see \fref{fig_Cs2}). For the
vibrational quantum number $v \sim 100$ the density of levels is
high due to unharmonicity and it is possible to find very close
levels of two different potential curves. This leads to enhanced
sensitivity to variation of $\mu$, as in \Eref{o}. Cold Cs$_2$
molecules can be produced in a particular quantum state by
photoassociation of Cs atoms in a trap.

Let us estimate the sensitivity of this experiment to variation of
$\alpha$ and $\mu$. For the electronic transition energy we can use
\Eref{hff2}. If we neglect unharmonicity, we can write the
transition frequency between close vibrational levels of the two
electronic terms in the form
\begin{align}\label{Cs2a}
 \omega=\omega_\mathrm{el,0}+qx
 +(v_2+\tfrac12)\,\omega_\mathrm{vib,2}
 -(v_1+\tfrac12)\,\omega_\mathrm{vib,1},
\end{align}
where $v_2\ll v_1$. The dependence of this frequency on constants is
given by:
\begin{align}\label{Cs2b}
 \delta\omega&\approx 2q\frac{\delta\alpha}{\alpha}
 -\frac{\omega_\mathrm{el,0}}{2}\frac{\delta\mu}{\mu}\,,
\end{align}
where we took into account that $\omega\ll\omega_\mathrm{el,0}$. A
very rough estimate of the factor $q$ can be done in the following
way. For the ground state of atomic Cs the $q$-factor is about
1100~cm$^{-1}$, which is close to $\tfrac14 \alpha^2 Z^2
\veps_{6s}$, where $\veps_{6s}$ is the ground-state binding energy.
If we assume that the same relation holds for the electronic
transition in molecule, we get $|q|\sim\tfrac14 \alpha^2
Z^2\omega_\mathrm{el,0}\sim 120$ cm$^{-1}$. Using this estimate and
\Eref{Cs2b} we get:
\begin{align}\label{Cs2c}
 \delta\omega&\approx -240\frac{\delta\alpha}{\alpha}
 -1600\frac{\delta\mu}{\mu}\,,
\end{align}
where we assume that relativistic corrections reduce dissociation
energy of the molecule, so $q$ is negative. This estimate shows that
the experiment with Cs$_2$ is mostly sensitive to variation of
$\mu$.

Estimate \eqref{Cs2c} is obtained in the harmonic approximation. As
mentioned above, for high vibrational states real potential is
highly unharmonic. This significantly decreases the sensitivity of
this experiment compared to the naive estimate \eqref{Cs2c}. It can
be easily seen either from the WKB approximation \cite{DeM04,DSS07},
or from an analytical solution for the Morse potential \cite{ZKY07}.
Quantization condition for vibrational spectrum in the WKB
approximation reads:
\begin{align}\label{Cs2d}
 \int_{R_1}^{R_2}\sqrt{2M(U(r)-E_n)}\,\mathrm{d}r
 &= \left(v+\tfrac12\right)\pi\,.
\end{align}
Differentiating this expression in $\mu$ we get:
\begin{align}\label{Cs2e}
 \delta E_v &=\frac{v+\tfrac12}{2\rho(E_v)} \frac{\delta\mu}{\mu}\,,
\end{align}
where $\rho(E_v)\equiv\left(\partial E_v/\partial
v\right)^{-1}\approx \left(E_v-E_{v-1}\right)^{-1}$ is the level
density. For the harmonic part of the potential,
$\rho=\mathrm{const}$ and the shift $\delta E_v$ grows linearly with
$v$, but for vibrational states near the dissociation limit the
level density $\rho(E) \longrightarrow \infty$ and $\delta E_v
\longrightarrow 0$. Consequently, maximum sensitivity $\sim 1000$
cm$^{-1}$ is reached at $v\approx 60$, and rapidly drops for higher
$v$. At present the group at Yale has found a conveniently close
vibrational level of the upper $^3\Sigma_u$ state for $v=138$, but
where the sensitivity is only $\sim 200$ cm$^{-1}$ \cite{DSS07}.
There are still good chances that there are other close levels with
smaller $v$, where the sensitivity may be several times higher.

It is important that because of unharmonicity, the sensitivity to
variation of $\alpha$ also decreases compared to the estimate
\eqref{Cs2c}. The reason for this is the following. For the highest
vibrational levels of the ground state, as well as for all levels of
the upper (weakly bound) state, the separation between nuclei is
large, $R\gtrsim 12$~a.u. (see \fref{fig_Cs2}). Thus, both
electronic wave functions are close to either symmetric (for
$^1\Sigma_g^+$) or antisymmetric combination (for $^3\Sigma_u^+$) of
atomic $6s$ functions:
\begin{align}
 \label{Cs2wf}
 \Psi_{g,u}(r_1,r_2) &\approx\frac{1}{\sqrt{2}}
 \left(6s^a(r_1) 6s^b(r_2) \pm 6s^b(r_1) 6s^a(r_2)\right).
\end{align}
Therefore, all relativistic corrections are (almost) the same for
both states.

Similar conclusions can be reached from the analysis of the Morse
potential:
\begin{align}
 \label{Cs2f}
 U_M(r)&=d\left(1-e^{-a(r-r_0)}\right)^2-d\,.
\end{align}
The eigenvalues for this potential are given by the analytical
expression:
\begin{align}
 \label{Cs2g}
 E_v&=\omega_0(v+\tfrac12)-\frac{\omega_0^2(v+\tfrac12)^2}{4d}-d\,,
\end{align}
where $\omega_0=2\pi a \sqrt{2d/M}$ and the last eigenvalue $E_N$ is
found from the conditions $E_{N+1}\le E_N$ and $E_{N-1}\le E_N$.
Obviously, $E_N$ is very close to zero and is practically
independent from any parameters of the model. Therefore, it is also
insensitive to variation of constants.

\begin{figure*}[htb]
 \includegraphics[scale=0.8]{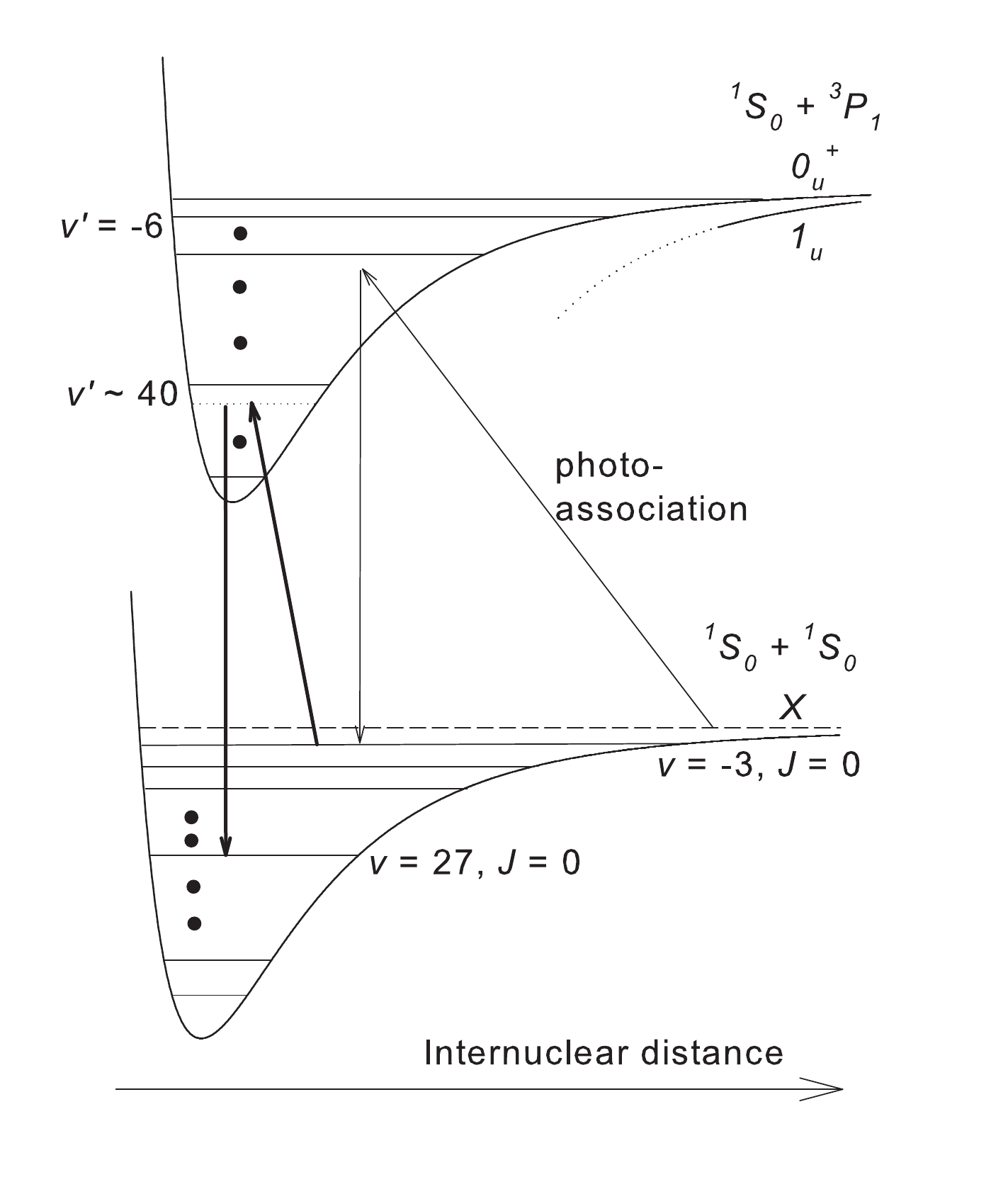}
 \caption{The scheme for Raman spectroscopy of Sr$_2$ ground-state
 vibrational spacings. A two-color photoassociation pulse prepares
 molecules in the $v=v_\mathrm{max}-2$ vibrational state (denoted on the plot
 as $v=-3$). Subsequently, a Raman pulse couples the $v=-3$ and $v=27$ states via
 $v'\approx 40$ level of the excited $0_u^+$ state (figure
 from Ref.~\cite{ZKY07}).} \label{fig_Sr2}
\end{figure*}

We see that highest absolute sensitivity is reached for vibrational
levels somewhere in the middle of the potential curve. However, in
this part of the spectrum there are no close levels of different
nature to maximize the relative sensitivity $\delta\omega/\omega$.
One can still use frequency combs to perform high-accuracy
measurements. This idea is used in the resent proposal by Zelevinsky
\etal\ \citet{ZKY07}, who suggest to use an optical lattice to trap
Sr$_2$. These molecules are formed by photoassociation in one of the
uppermost vibrational levels of the ground electronic states (see
\fref{fig_Sr2}). As we saw above, this level is not sensitive to the
variation of $\mu$. At the next stage, a Raman transition is driven
to one of the most sensitive levels in the middle of the potential
well. This way it is possible to get the highest possible absolute
sensitivity for a given molecule. Unfortunately, the dissociation
energy for Sr$_2$ is only about 1000 cm$^{-1}$, which is 3 times
smaller than for Cs$_2$. Because of this, the highest sensitivity
for the Sr$_2$ molecule is about 270 cm$^{-1}$, i.e. only slightly
higher than the sensitivity of the $v=138$ level in Cs$_2$.
Therefore, it may be useful to try to apply this scheme to some
other molecule with larger dissociation energy. Note that in the
experiment with Sr$_2$, the sensitivity to $\alpha$-variation is
additionally suppressed by a factor $(38/55)^2\approx 1/2$ because
of the smaller $Z$.

\section{Experiments with hydrogen molecular ions H$_2^+$ and
HD$^+$}\label{H2}

Hydrogen molecular ions are very attractive for fundamental studies
because of their theoretical simplicity and experimental possibility
of their cooling and trapping. Using H$_2^+$ and HD$^+$ ions for
studying time-variation of electron-to-proton and proton-to-deuteron
mass ratios $\mu=m_e/m_p$ and $m_p/m_d$ has been suggested in
Refs.~\cite{FRA04,SK05}. Because of the unharmonicity, the ratio of
the two vibrational transitions with very different vibrational
quantum numbers is $\mu$-dependent \cite{FRA04}. There is no
enhancement of the relative effect here, but the lines are very
narrow and high-precision measurements are possible using frequency
combs.

Recently HD$^+$ ion has been cooled to 50 mK and trapped in a linear
rf trap \cite{KRW07}. This allowed to measure the rovibrational
transition $v,N=0,2 \rtw v',N'=4,3$ with an absolute accuracy of 0.5
MHz. Using sensitivity coefficient from \cite{SK05} one can see,
that this accuracy translates in to $5\times 10^{-9}$ (5 ppb)
accuracy for $\mu$. Note that modern molecular theory of HD$^+$ has
comparable accuracy \cite{Kor06}. Thus, a direct comparison between
theory and experiment allows to determine the absolute value of
$\mu$ to 5 ppb.

\section{Conclusions}\label{concl}

We have seen that both diatomic and polyatomic molecules are used in
astrophysics to study possible variation of the electron-to-proton
mass ratio $\mu$ on a time scale from 6 to 12 billion years. Results
of these studies are inconclusive, see Eqs.~\eqref{ubach_result1},
\eqref{OH1}, and \eqref{nh3final}. The situation is similar for the
astrophysical search for $\alpha$-variation. In principle, all these
results can be explained by complex evolution of $\mu$ and $\alpha$
in space and time. Or, more likely, there are some systematic
errors, which are not fully understood. Therefore it is extremely
important to supplement astrophysical studies with laboratory
measurements of present-day variation of these constants. This work
is currently going on in many groups. Most of them use atomic
frequency standards and atomic clocks. In this chapter we discussed
several resent ideas and proposals on how to increase the
sensitivity of laboratory tests by using molecules instead of atoms.

The only molecular experiment \cite{Cha07,AGG05}, which has reached
the stage of placing the limit on the time-variation of fundamental
constants \eqref{SF6a}, used a supersonic molecular beam of SF$_6$.
Even though this experiment is less sensitive than the best atomic
experiments, it constrains a different combination of fundamental
constants. This allows to combine it with the results of
atomic-clock experiments \cite{Peik2004,Fischer2004,clock_1} to
place the most stringent laboratory limit \eqref{SF6b} on
time-variation of $\mu$. The linewidth in this experiment,
$\Gamma\approx 200$~Hz, was determined by the time-of-flight through
the 1-m Ramsey interferometer. A similar problem with the linewidth
has prevented the use of the ND$_3$ beam to perform competitive
experiment on time-variation \cite{VKB04}. Using cold molecules
would allow to reduce the linewidth by several orders of magnitude
and drastically raise the sensitivity of molecular experiments.

We have seen that for such diatomic radicals as Cl$_2^+$ and SiBr
there are narrow levels of different nature separated by intervals
$\lesssim 1$~cm$^{-1}$. The natural widths of these levels are on
the order of $10^{-2}~\mathrm{Hz}$. This is comparable to the
accuracy necessary to reach the sensitivity of $\delta \alpha /
\alpha \sim 10^{-15}$, similar to that of the best modern laboratory
tests. In the high-precision frequency measurements, the measurement
accuracy is typically few orders of magnitude better than the
linewidth. Of course, in order to benefit from such narrow lines, it
is crucial to be able to cool and trap the molecules. In this
respect the ion Cl$_2^+$ looks more promising.

Even higher sensitivity to the temporal variation of $\alpha$ can be
found in HfF$^+$ and similar molecular ions, which are being
considered for the search of the electron EDM at JILA
\cite{SC04,MBD06,PMI06}. The transition amplitude between
$^3\Delta_1$ and $^1\Sigma_0$ of HfF$^+$ ion is also suppressed. The
transition width is larger than for Cl$_2^+$ and SiBr because of the
larger value of $Z$ and higher frequency $\omega_f$. In
Ref.~\cite{PMI06} the width of $^3\Delta_1$ state was estimated to
be about 2~Hz. This width is also of the same order of magnitude as
the expected frequency shift for $\delta \alpha / \alpha \sim
10^{-15}$. At present not much is known about these molecular ions.
More spectroscopic and theoretical data are needed to estimate the
sensitivity to $\alpha$-variation reliably. We hope that this review
may stimulate further studies in this direction. Additional
advantage here is the possibility to measure electron EDM and
$\alpha$-variation using the same molecule and a similar
experimental setup.

Preliminary spectroscopic experiment with the Cs$_2$ molecule has
been recently finished at Yale \cite{DSS07}. The electron transition
in Cs$_2$ goes between the $^3\Sigma_u^+$ and $^1\Sigma_g^-$ states,
and, to a first approximation, is independent of $\alpha$. On the
other hand the sensitivity to $\mu$ may be enhanced because of the
large number of vibrational quanta needed to match the electronic
transition. However, the unharmonicity of the potential curve near
the dissociation limit suppresses this enhancement for very high
vibrational levels. As a result, the sensitivity to variation of
$\mu$ for the $v=138$ level is about the same as in \Eref{do1}. It
is possible that there are other close levels with smaller
vibrational quantum number $v$ and, consequently, with higher
sensitivity. Even if such levels are not found, the experiment with
the $v=138$ level may improve present limit on variation of $\mu$ by
several orders of magnitude.

An experiment with the Sr$_2$ molecule was recently proposed at JILA
\cite{ZKY07}. This experiment potentially has similar sensitivity to
variation of $\mu$ as the experiment with Cs$_2$ and both of them
are complementary to the experiments with molecular radicals, which
are mostly sensitive to $\alpha$-variation \cite{FK2}.

Finally, we have seen that the inversion spectra of such polyatomic
molecules as NH$_3$ and ND$_3$ are potentially even more sensitive
to variation of $\mu$. This has already been used in astrophysics to
place the most stringent limit \eqref{nh3final} on the
time-variation of $\mu$ on the cosmological timescale. Corresponding
laboratory experiments require very slow molecular beams, fountains,
or molecular traps. The work in this direction is going on
\cite{VKB04}.

To conclude this chapter, we see that this field is rapidly
developing and new interesting results can be expected in the near
future.

\begin{acknowledgments}
We want to thank J.\ Ye, D.\ DeMille, and S.\ Schiller for their
extremely useful comments. We are particularly grateful to D.\
Budker, whose advise significantly improved the whole manuscript.
\end{acknowledgments}

%\bibliographystyle{apsrev}%
%\bibliography{../bib/alpha,../bib/julia_w,../bib/my_ref_w}
%\end{document}

%%%%  >>>>eoref<<<<<
%%%%  >>>>eoref<<<<<
%%%%  >>>>eoref<<<<<
%%%%  >>>>eoref<<<<<

\end{document}